# Three Remarks On Asset Pricing

Victor Olkhov

Independent, Moscow, Russia

victor.olkhov@gmail.com

ORCID: 0000-0003-0944-5113

**Abstract**

We consider the consumption-based asset-pricing model, derive a new modified basic pricing equation, and present its successive approximations using the Taylor series expansions of the investor's utility during the averaging time interval. For linear and quadratic Taylor approximations, we derive new expressions for the mean price, mean payoff, volatility, skewness, and the asset's amount that define the maximum of the investor's utility. We discuss the market-based origin of price probability. We use volume weighted average price (VWAP) as a market-based average price and introduce market-based price volatility. The use of VWAP results in zero correlations between the price *p* and trade volume *U*. We derive a correlation between price *p* and squares of trade volume $U^2$ and between squares of price $p^2$ and volume $U^2$. To predict market-based price volatility, one should forecast the *2*-d statistical moments of the market trade values and volumes at the same horizon *T*.

---

This research received no support, specific grants, or financial assistance from funding agencies in the public, commercial, or nonprofit sectors. We welcome valuable offers of grants, support and positions.

## 1. Introduction

In this paper, we discuss the consumption-based model and show how the few remarks generated by the reality of market trade could impact the performance of asset pricing.

The literature on asset pricing is huge and boundless. We refer only a tiny part of the endless publications on asset pricing, starting with CAPM by William Sharpe (1964), which was followed by various modifications such as Intertemporal CAPM by Robert Merton (1973), the Arbitrage Theory of Capital Asset Pricing by Stephen Ross (1976), the consumption-based asset pricing model described by Darrell Duffie and William Zame (1989), John Cochrane (2001), John Campbell (2002), and many others. Cochrane (2001) demonstrates that the consumption-based asset-pricing framework provides unified approach for most variations of pricing models. We consider Cochrane's description of the consumption-based model as a certain unification of current pricing theories and chose his monograph (Cochrane, 2001) as our main reference. If one accepts Cochrane's assertion that the consumption-based model and the basic pricing equation describe the results of most pricing theories, then our remarks, approximations, and results make sense for other asset pricing theories.

Actually, economic and financial models describe approximations that capture certain averaging, smoothness, and coarsening of the economic reality. Our first remark concerns the importance of using a particular averaging interval $\Delta$ as a determining factor in asset pricing models. Indeed, current stock markets support initial time axis division determined by the time series of the trades performed at moments $t_i$ with a time shift $\varepsilon = t_i - t_{i-1}$ between trades. For simplicity, we consider the time shift $\varepsilon$ as a constant. As usual, the time shift $\varepsilon$ is sufficiently small and can be equal to a second or even a fraction of a second. That is not too useful to model asset prices at a horizon $T$ that can be equal to a month, quarter, or year. However, records of trade time series with time shifts $\varepsilon$ determine the initial market time axis division and define the discrete nature of market trade data. To evaluate any reasonable pricing model at a given horizon $T$ one should choose the averaging scale $\Delta$, which should obey $\varepsilon << \Delta < T$. The choice of the averaging interval $\Delta$ determines the scale of averaging of the initial trade time series and provides the transition from the initial market time division multiple of $\varepsilon$ to the averaged time division multiple of $\Delta$. Below, we show how the choice of the averaging interval results in a modification of the consumption-based utility function and the basic pricing equation.

Our second remark indicates that the choice of the averaging interval $\Delta$ permits us to expand the utility function and the basic pricing equation into Taylor series near the average values of price and payoff and then average the fluctuating terms of the series. Mathematical expectations of linear and quadratic Taylor approximations of the basic pricing equation by price and payoff variations during *Δ* give new expressions of the mean price and payoff, their volatilities, skewness, and other factors. Actually, even a linear Taylor expansion of the basic pricing equation demonstrates that the famous statement "price equals expected discounted payoff," with which Cochrane (2001) and Markus Brunnermeier (2015) begin their papers, describes approximations for cases with zero price volatility “today” and "next day". As we show in Sec. 4, the Taylor expansion of the modified basic pricing equation determines relations between mean price and price volatility during the current period and mean payoff and payoff volatility during the "next day" period. Further in Section 4, we derive relations that extend the results of the consumption-based model.

Our third remark concerns the market origin of the asset price probability as the major issue of any pricing model and financial economics as a whole. We consider the randomness of market trade values and volumes as the origin of price stochasticity. We take volume weighted average price (VWAP) that was introduced by Berkowitz, et al., (1988) as the 1-st statistical moment of the market-based price probability and introduce the 2-d statistical moment and price volatility. We show that the use of VWAP results in zero correlations between price $p$ and trade volume $U$. We derive correlations between price $p$ and squares of trade volume $U^2$ and correlations between squares of price $p^2$ and squares of volume $U^2$.

We propose that readers become familiar with Cochrane (2001) and refer to his monograph for any clarifications. In Sec. 2, we briefly recall the main notions of asset pricing according to Cochrane (2001). In Sec. 3, we consider remarks on the averaging interval *Δ* and explain the necessity for modification of the basic pricing equation. In Sec. 4, we discuss the Taylor series expansion of the utility functions and derive successive approximations of the modified basic pricing equation in linear and quadratic approximations by the price and payoff variations. In Sec. 5, we introduce the 1-st and 2-d market-based price statistical moments and price volatility. In Sec. 6, we consider how market-based price statistical moments define price-volume correlations. The conclusion is in Sec.7. In App. A, we present calculations that define the maximum of an investor’s utility.

Equation (4.5) means equation 5 in Sec. 4, and (A.2) notes equation 2 in Appendix A. We assume that readers are familiar with the basic notions of probability, statistical moments, etc.

## 2. Brief Notations

In this section, we follow Cochrane (2001) and briefly present its main notations for asset pricing. The consumption-based basic pricing equation takes form:

$$p = E[m\,x] \tag{2.1}$$

In (2.1), $p$ denotes the asset price at date $t$, $x=p_{t+1}+d_{t+1}$ – payoff, $p_{t+1}$ - price and $d_{t+1}$ - dividends at date $t+1$, $m$ - stochastic discount factor, and $E[..]$ – mathematical expectation at day $t+1$ made by the forecast under the information available at date $t$. Cochrane (2001) considers equation (2.1) in various forms to show that most asset pricing models can be described by similar equations. For convenience, we briefly reproduce the derivation of the basic pricing equation (2.1). Cochrane models investors by a utility function $W(c_t;\, c_{t+1})$ defined over current $c_t$ and future $c_{t+1}$ values of consumption at dates $t$ and $t+1$ respectively.

$$W(c_t;\, c_{t+1}) = w(c_t) + \beta E[w(c_{t+1})] \tag{2.2}$$

$$c_t = e_t - p\xi \quad ; \quad c_{t+1} = e_{t+1} + x\xi \tag{2.3}$$

$$x = p_{t+1} + d_{t+1} \tag{2.4}$$

In (2.2), $w(c_t)$ and $w(c_{t+1})$ are utility functions at dates $t$ and $t+1$; in (2.3), $e_t$ and $e_{t+1}$ "denote the original consumption level (if the investor bought none of the asset), and $\xi$ denotes the amount of the asset he chooses to buy" (Cochrane, 2001). Cochrane calls $\beta$ a "subjective discount factor that captures impatience of future consumption." The first-order maximum condition for (2.2) by the amount of assets $\xi$ is fulfilled by putting the derivative of (2.2) by $\xi$ equals zero:

$$max_{\xi}\, W(c_t;\, c_{t+1}) \;\leftrightarrow\; \frac{\partial}{\partial \xi} W(c_t;\, c_{t+1}) = 0 \tag{2.5}$$

From (2.2-2.5), obtain:

$$p = \beta E\left[\frac{w'(c_{t+1})}{w'(c_t)}\, x\right] = E[mx] \quad ; \quad m = \beta \frac{w'(c_{t+1})}{w'(c_t)} \quad ; \quad w'(c) \equiv \frac{d}{dc} w(c) \tag{2.6}$$

Equations (2.6) reproduce (2.1) for $m$ (2.6). We refer to Cochrane (2001) for any details.

## 3. Remarks on Time Scales

We start with simple remarks on the averaging of economic and financial time series. Any economic or financial model, and asset pricing in particular, approximates real processes by averaging them over a certain time interval $\Delta$. To describe asset pricing, one should take into account that market trade time series are the only source of price variations. The interval $\varepsilon$ between market transactions can be very small and can be equal to a second or even a fraction of a second. Initial records of price time series $p(t_i)$ with time-shift $\varepsilon$ are very irregular and not very useful for modeling and forecasting asset prices at any reasonable time horizon $T$ that can be equal to a week, month, year, etc. To derive a reasonable description of asset prices, one should choose an averaging interval $\Delta$ and smooth variations of market prices during $\Delta$. The choice of an averaging interval $\Delta$ is a very important challenge for each investor. The choice of a long interval $\Delta$, which equals weeks or months, would result in smooth dynamics and stable predictions of the averaged variables but would limit the capacity to take investment decisions "this hour" or "today." Short averaging interval $\Delta$, such as hours or days, improve the ability to make "this hour" decisions, but average variables could be under the impact of multiple perturbations with periods equal to days or weeks. Different averaging intervals cause different random properties of variables, and different models describe averaged variables.

To perform a transition from the initial market trade time axis division, which is a multiple of $\varepsilon$ one should choose a time interval $\Delta$ such as $\varepsilon<<\Delta<T$ and an average price time series $p(t_i)$ during $\Delta$. The time shift $\Delta = t(k) - t(k-1)$ introduces a new time axis division multiple of $\Delta$. One can consider averaging intervals $\Delta_k$ as (3.1):

$$\Delta_k = \left[\, t(k) - \frac{\Delta}{2}\, ; t(k) + \frac{\Delta}{2} \right] \;\; ; \quad t(k) = t(0) + k\,\Delta \;\; ; \quad k = 0,\ 1, 2, .. \qquad (3.1)$$

We take the duration of each averaging interval $\Delta_k$ equal $\Delta$. One can consider time $t=t(0)$ as the moment "today" and the "next day" at time $t+1$ as $t(K)$ for some K>>1. What is most important is that the time axis division "today" at $t$ and the "next-day" at $t+1$ must be the same. Indeed, time axis divisions can't be measured "today" in hours and "next-day" in weeks. Utility (2.2) "today" at moment $t$ and "next-day" at $t+1$ should have the same time axis divisions. Averaging any time series at the "next-day" at $t+1$ during the interval $\Delta$ undoubtedly implies averaging "today" at date $t$ during an equal time interval $\Delta$ and vice versa. Thus, if the utility (2.2) is averaged at $t+1$ during the interval $\Delta$, then the utility (2.2) also should be averaged at date $t$ during the same interval $\Delta$ and (2.2) should take the form:

$$W(c_t;\ c_{t+1}) = E_t[w(c_t)] + \beta E[w(c_{t+1})] \qquad (3.2)$$

In (3.2), we denote $E_t[..]$ a mathematical expectation "today" at date $t$ during $\Delta$. It does not matter how one considers the market price time series "today" – as a random or as irregular. Mathematical expectation $E_t[..]$ performs smoothing of the random or irregular time series by aggregating data during $\Delta$ under a particular probability measure. Mathematical expectations $E_t[..]$ at $t$ and $E[..]$ at $t+1$ during the same averaging intervals $\Delta$ establish identical time division of the problem at dates $t$ and $t+1$ in (3.2). Hence, relations similar to (2.5; 2.6) should cause a modification of the basic pricing equation (2.1; 2.6) in the form (3.3):

$$E_t[p\; w'(c_t)] = \beta E\; [x\; w'(c_{t+1})] \qquad (3.3)$$

Cochrane (2001) takes the "subjective discount factor" $\beta$ as non-random, and we do the same. Mathematical expectation $E_t[..]$ averages $pw'(c_t)$ over random price $p$ fluctuations during $\Delta$

"today". On the right side, $E[xw'(c_{t+1})]$ averages $xw'(c_{t+1})$ over random payoff fluctuations during $\Delta$ "next day" on the basis of data available at date $t$ "today."

## 4. Remarks on Taylor series

Relation (2.5) presents the first-order condition for the amount of assets $\xi_{max}$ that brings maximum of the investor's utility (2.2) or (3.2). Let us choose the averaging interval $\Delta$ and take the price $p$ at date $t$ and the payoff $x$ at date $t+1$ during $\Delta$ as:

$$p = p_0 + \delta p \; ; \quad x = x_0 + \delta x \; ; \quad E_t[p] = p_0 \; ; \; E[x] = x_0 \tag{4.1}$$

$$E_t[\delta p] = E[\delta x] = 0 \; ; \; \sigma^2(p) = E_t[\delta^2 p] \; ; \; \sigma^2(x) = E[\delta^2 x] \tag{4.2}$$

Relations (4.1; 4.2) denote the average price $p_0$ and its volatility $\sigma^2(p)$ at date $t$ and the average payoff $x_0$ and its volatility $\sigma^2(x)$ at date $t+1$. We consider $\delta p$ and $\delta x$ as random fluctuations of price and payoff during $\Delta$. We highlight that we consider averaging during $\Delta$ as averaging of a random variable or as smoothing of an irregular variable. Thus, $E_t[p]$ – at date $t$ smooths the random or irregular price $p$ (4.1) during $\Delta$ and $E[x]$ – averages the random payoff $x$ (4.1) during $\Delta$ at date $t+1$. We present the derivatives of utility functions in (3.3) by Taylor series in a linear approximation by $\delta p$ and $\delta x$ during $\Delta$:

$$w'(c_t) = w'(c_{t;0}) - \xi w''(c_{t;0})\delta p \quad ; \quad w'(c_{t+1}) = w'(c_{t+1;0}) + \xi w''(c_{t+1;0})\delta x \tag{4.3}$$

$$c_{t;0} = e_t - p_0 \xi \quad ; \quad c_{t+1;0} = e_{t+1} + x_0 \xi$$

Now substitute (4.3) into (3.3), and due to (4.2), obtain the equation (4.4):

$$w'(c_{t;0})p_0 - \xi w''(c_{t;0})\sigma^2(p) = \beta w'(c_{t+1;0})x_0 + \beta \xi w''(c_{t+1;0})\sigma^2(x) \tag{4.4}$$

Taylor series are simple mathematical tools, and Cochrane (2001) also used them. We underline: Taylor series and (4.1-4.4) are determined by the duration of $\Delta$. The change of $\Delta$ can implies a change of the mean price $p_0$, the mean payoff $x_0$ and their volatilities $\sigma^2(p)$, $\sigma^2(x)$ (4.2). Equation (4.4) is a linear approximation of the price and payoff fluctuations of the first-order max conditions (2.5) and assesses the root $\xi_{max}$ that brings the maximum of the utility $W(c_t;c_{t+1})$ (3.2):

$$\xi_{max} = \frac{w'(c_{t;0})p_0 - \beta w'(c_{t+1;0})x_0}{w''(c_{t;0})\sigma^2(p) + \beta w''(c_{t+1;0})\,\sigma^2(x)} \tag{4.5}$$

We note that (4.5) is not an "exact" solution for $\xi_{max}$ as derivatives of utilities $w'$ and $w''$ also depend on $\xi_{max}$ as it follows from (4.3). However, (4.5) gives an assessment of $\xi_{max}$ in a linear approximation by Taylor series $\delta p$ and $\delta x$ averaged during $\Delta$. Let us highlight that the $\xi_{max}$ (4.5) depends on the price volatility $\sigma^2(p)$ at date $t$ and on the forecast of payoff volatility $\sigma^2(x)$ at date $t+1$ (4.2).

It is clear that sequential iterations may give more accurate approximations of $\xi_{max}$. Nevertheless, our approach and (4.5) give a new look at the basic equation (2.6; 3.3). If one follows the standard derivation of (2.6) (Cochrane, 2001) and neglects the averaging at date $t$ in the left side (3.3), then (2.6; 4.5) give

$$\xi_{max} = \frac{w'(c_t)p - \beta w'(c_{t+1;0})x_0}{\beta w''(c_{t+1;0})\sigma^2(x)} \tag{4.6}$$

Relations (4.6) show that even the standard form of the basic equation (2.6) hides the dependence of the amount of assets $\xi_{max}$ on the payoff volatility $\sigma^2(x)$ at date $t+1$. If one has an independent assessment of $\xi_{max}$ then one can present (4.6) in a way similar to the basic equation (2.6):

$$p = \frac{w'(c_{t+1;0})}{w'(c_t)}\beta x_0 + \xi_{max}\frac{w''(c_{t+1;0})}{w'(c_t)}\beta\sigma^2(x) \tag{4.7}$$

Otherwise, if there are no independent assessments of $\xi_{max}$, then one should consider (4.6) as the solution of the first order maximum condition (2.5), which presents the root $\xi_{max}$ of the amount of assets, determined for the given values in the right hand of (4.6). In that case, the basic pricing equations (2.1; 2.6; 4.7) make almost no sense, as the value of $\xi_{max}$ in (4.7) is not determined. We consider this misstep – using the maximum condition (2.5) to determine the basic pricing equation (2.1; 4.7) instead of defining $\xi_{max}$ as a root of the maximum condition (2.5) - a significant oversight of the consumption-based asset pricing model, which requires essential clarifications. One can transform (4.7) similar to (2.6):

$$p = m_0 x_0 + \xi_{max} m_1 \sigma^2(x) \tag{4.8}$$

$$m_0 = \frac{w'(c_{t+1;0})}{w'(c_t)}\beta \;\; ; \;\; m_1 = \frac{w''(c_{t+1;0})}{w'(c_t)}\beta \tag{4.9}$$

For the given $\xi_{max}$ equation (4.8) in a linear approximation by Taylor series describes the dependence of the price $p$ at date $t$ (3.1) on the mean discount factors $m_0$ and $m_1$ (4.9), the mean payoff $x_0$ (4.1), and the payoff volatility $\sigma^2(x)$ during $\Delta$. Let us stress that while the mean discount factor $m_0>0$, the mean discount factor $m_1<0$ because utility $w'(c_t)>0$ and $w''(c_t)<0$ for all $t$. Hence, irremovable payoff volatility $\sigma^2(x)$ at day $t+1$ states that price $p$ at day $t$ always less than discounted mean payoff $x_0$:

$$p < m_0 x_0 \quad ; \quad \xi_{max} m_1 \sigma^2(x) < 0$$

One can consider (4.8) as a linear Taylor expansion of (2.1; 2.6). However, equation (4.4) presents the dependence of mean price $p_0$ at day $t$ on price volatility $\sigma^2(p)$ at day $t$, mean payoff $x_0$ and payoff volatility $\sigma^2(x)$ at day $t+1$. That definitely enlarges the conventional statement that "price equals expected discounted payoff". We indicate that (4.6-4.9) makes sense for the given value of $\xi_{max}$. As the price $p$ in (4.8) should be positive, hence $\xi_{max}$ should obey inequality (4.10):

$$0 < \xi_{max} < -\frac{w'(c_{t+1;0})}{w''(c_{t+1;0})}\frac{x_0}{\sigma^2(x)} \tag{4.10}$$

For the conventional power utility (A.2) (Cochrane, 2001), from (4.3) obtain for (4.10):

$$w(c) = \frac{1}{1-\alpha}c^{1-\alpha} \quad ; \quad \frac{w'(c)}{w''(c)} = -\frac{c}{\alpha} \quad ; \quad 0 < \alpha \leq 1$$

inequality (4.10) valid always if

$$\alpha\, \sigma^2(x) < x_0^2$$

For this approximation (4.10) limits the value of $\xi_{max}$. For (4.4; 4.5) obtain equations similar to (4.8; 4.9):

$$m_0 = \frac{w'(c_{t+1;0})}{w'(c_{t;0})}\beta > 0\, ; \;\; m_1 = \frac{w''(c_{t+1;0})}{w'(c_{t;0})}\beta < 0\, ; \;\; m_2 = \frac{w''(c_{t;0})}{w'(c_{t;0})} < 0 \tag{4.11}$$

$$p_0 = m_0 x_0 + \xi_{max}[m_1\sigma^2(x) + m_2\sigma^2(p)] \tag{4.12}$$

We use the same notions $m_0$, $m_1$ to denote the discount factors, taking into account the replacement of $w'(c_t)$ in (4.9) by $w'(c_t;0)$ in (4.11; 4.12). The modified basic pricing equation (4.12) at date $t$ describes the dependence of the mean price $p_0$ on the price volatility $\sigma^2(p)$ at date $t$, the mean payoff $x_0$ and the payoff volatility $\sigma^2(x)$ at date $t+1$ averaged during $\Delta$.

Equation (4.12) illustrates the well-known practice that high price volatility $\sigma^2(p)$ at date $t$ and a forecast of high payoff volatility $\sigma^2(x)$ at date $t+1$ may cause a decline in the mean price $p_0$ at date $t$.

### *4.1 The Idiosyncratic Risk*

Here we follow (Cochrane, 2001) and consider the usage of the Taylor series for his example of the idiosyncratic risk for which the payoff $x$ in (2.6) is not correlated with the discount factor $m$ at moment $t+1$:

$$cov(m, x) = 0 \tag{4.13}$$

In this case equation (2.6) takes form:

$$p = E[mx] = E[m]E[x] + cov(m, x) = E[m]x_0 = \frac{x_0}{R_f} \tag{4.14}$$

The risk-free rate $R_f$ in (4.14) is known ahead (Cochrane, 2001). From (4.3) in a linear approximation by $\delta x$ Taylor series for the derivative of the utility $w'(c_{t+1})$:

$$w'(c_{t+1}) = w'\big(c_{t+1;0}\big) + w''\big(c_{t+1;0}\big)\xi\delta x \tag{4.15}$$

Hence, the discount factor $m$ (2.6) takes form:

$$m = \beta\frac{w'(c_{t+1})}{w'(c_t)} = \frac{\beta}{w'(c_t)}\left[w'\big(c_{t+1;0}\big) + w''\big(c_{t+1;0}\big)\xi\delta x\right]$$

$$E[m] = \bar{m} = \beta \frac{w'(c_{t+1;0})}{w'(c_t)} \quad ; \quad \beta E\left[\frac{w'(c_{t+1})}{w'(c_t)}\right] x_0 = \frac{x_0}{R_f} \quad ; \quad E[w'(c_{t+1})x] = 0$$

$$\delta m = m - \bar{m} = \frac{\beta}{w'(c_t)} w''\left(c_{t+1;0}\right)\xi \delta x$$

Hence, (4.13) implies:

$$cov(m, x) = E[\delta m \delta x] = \beta \frac{w''(c_{t+1;0})}{w'(c_t)} \xi_{max} \sigma^2(x) = 0 \qquad (4.16)$$

That results in zero payoff volatility $\sigma^2(x)=0$. Of course, zero payoff volatility does not model market reality, but (4.16) reflects the restrictions of the linear approximation (4.15). To overcome this discrepancy, take into account the Taylor series up to the second power by $\delta^2 x$:

$$w'(c_{t+1}) = w'\left(c_{t+1;0}\right) + w''\left(c_{t+1;0}\right)\xi\delta x + \frac{1}{2} w'''\left(c_{t+1;0}\right)\xi^2 \delta^2 x \qquad (4.17)$$

$$m = \beta \frac{w'(c_{t+1})}{w'(c_t)} = \frac{\beta}{w'(c_t)}\left[w'\left(c_{t+1;0}\right) + w''\left(c_{t+1;0}\right)\xi\delta x + \frac{1}{2} w'''\left(c_{t+1;0}\right)\xi^2 \delta^2 x\right] \qquad (4.18)$$

For this case, the mean discount factor $E[m]$ takes the form:

$$E[m] = \bar{m} = \frac{\beta}{w'(c_t)}\left[w'\left(c_{t+1;0}\right) + \frac{1}{2} w'''\left(c_{t+1;0}\right)\xi^2 \sigma^2(x)\right] \qquad (4.19)$$

and variations of the discount factor $\delta m$:

$$\delta m = m - \bar{m} = \frac{\beta}{w'(c_t)}\ \left[w''\left(c_{t+1;0}\right)\xi\delta x + \frac{1}{2} w'''\left(c_{t+1;0}\right)\xi^2\{\delta^2 x - \sigma^2(x)\}\right.$$

Thus the Taylor series approximation up to the second power by $\delta^2 x$ gives:

$$cov(m, x) = E[\delta m \delta x] = \left[w''\left(c_{t+1;0}\right)\xi\sigma^2(x) + \frac{1}{2} w'''\left(c_{t+1;0}\right)\xi^2\, \gamma^3(x)\right] = 0 \qquad (4.20)$$

$$\gamma^3(x) = E[\delta^3 x] \quad ; \quad Sk(x) = \frac{\gamma^3(x)}{\sigma^3(x)} \qquad (4.21)$$

*Sk(x)* – denotes normalized payoff skewness at date *t+1,* treated as the measure of asymmetry of the probability distribution during $\Delta$. For approximation (4.18) from (4.20; 4.21), obtain relations on the skewness *Sk(x)* and $\xi_{max}$:

$$\xi_{max}\, Sk(x)\sigma(x) = -2 \frac{w''(c_{t+1;0})}{w'''(c_{t+1;0})} \qquad (4.22)$$

For the conventional power utility (A.2)

$$w(c) = \frac{1}{1-\alpha} c^{1-\alpha}$$

and (4.3) relations (4.22) take the form

$$\xi_{max} = \frac{2e_{t+1}}{(1+\alpha)Sk(x)\sigma(x) - 2x_0} \qquad (4.23)$$

It is assumed that the second derivative of utility $w''(c_{t+1})<0$ always negative and the third derivative $w'''(c_{t+1})>0$ is positive, and hence the right side in (4.22) is positive. Hence, to get a positive $\xi_{max}$ for (4.23) for the power utility (A.2), the payoff skewness *Sk(x)* should obey inequality (4.24) that defines the lower limit of the payoff skewness *Sk(x):*

$$Sk(x) > \frac{2x_0}{(1+\alpha)\sigma(x)} \tag{4.24}$$

In (4.14), $R_f$ denotes the risk-free rate. Hence, (4.19; 4.22; 4.24) define relations:

$$\frac{\beta}{w'(c_t)}\left[w'\left(c_{t+1;0}\right) + \frac{1}{2}w'''\left(c_{t+1;0}\right)\xi_{max}{}^2\sigma^2(x)\right] = \frac{1}{R_f}$$

$$\frac{1}{2}\xi_{max}{}^2\sigma^2(x) = \frac{1}{\beta R_f}\frac{w'(c_t)}{w'''(c_{t+1;0})} - \frac{w'(c_{t+1;0})}{w'''(c_{t+1;0})}$$

$$Sk^2(x) = \frac{R_f}{1-m_0R_f}\frac{m_1^2}{m_3} > \frac{4x_0^2}{(1+\alpha)^2\sigma^2(x)} \quad ; \quad m_0 < 1/R_f$$

$$\frac{\sigma^2(x)}{4x_0^2} > \frac{m_3}{m_1^2}\frac{1-m_0R_f}{(1+\alpha)^2R_f} \tag{4.25}$$

Inequality (4.25) establishes the lower limit on the payoff volatility $\sigma^2(x)$ normalized by the square of the mean payoff $x_0^2$. The lower limit on the right side of (4.25) is determined by the discount factors (4.26), the risk-free rate $R_f$, and the conventional power utility factor $\alpha$ (A.2).

$$m_0 = \beta\frac{w'(c_{t+1;0})}{w'(c_t)} \; ; \; m_1 = \beta\frac{w''(c_{t+1;0})}{w'(c_t)} \; ; \; m_3 = \beta\frac{w'''(c_{t+1;0})}{w'(c_t)} \tag{4.26}$$

The coefficients in (4.26) differ a little from (4.1), as (4.26) takes the denominator $w'(c_t)$ instead of $w'(c_{t;0})$ in (4.11), but we use the same letters to avoid extra notations. The similar calculations for (3.2; 3.3) describe both the price volatility $\sigma^2(p)$ and price skewness $Sk(p)$ at date $t$ and the payoff volatility $\sigma^2(x)$ and payoff skewness $Sk(x)$ at date $t+1$. Further approximations by the Taylor series of the utility derivative $w'(c_t)$ up to $\delta^3p$ and $w'(c_{t+1})$ up to $\delta^3x$ similar to (4.17) give assessments of kurtosis of the price probability at date $t$ and the kurtosis of the payoff probability at date $t+1$ estimated during interval $\Delta$, but we omit these assessments for brevity.

### *4.2 The Utility Maximum*

Relations (2.5) define the first-order condition of maximum of the utility $W(c_t;c_{t+1})$ (2.2; 3.2). To confirm that function $W(c_t;c_{t+1})$ has a maximum at $\xi_{max}$, the first order condition (2.5) must be supplemented by the 2-d order condition:

$$\frac{\partial^2}{\partial\xi^2}W(c_t;c_{t+1}) < 0 \tag{4.27}$$

The use of (4.27) has interesting consequences. From (2.2–2.4) and (4.27), obtain:

$$p^2 > -\frac{\beta}{w''(c_t)}E[x^2\, w''(c_{t+1})\,] \tag{4.28}$$

The linear Taylor series expansion of the second derivative of the utility $w''(c_{t+1})$ by $\delta x$ give:

$$w''(c_{t+1}) = w''\left(c_{t+1;0}\right) + w'''\left(c_{t+1;0}\right)\xi\delta x$$

Then (4.28) takes the form:

$$p^2 > -\beta\frac{w''(c_{t+1;0})}{w''(c_t)}[x_0{}^2 + \sigma^2(x)] - \beta\frac{w'''(c_{t+1;0})}{w''(c_t)}\xi_{max}\,[2x_0\sigma^2(x) + \gamma^3(x)] \tag{4.29}$$

For the power utility (A.2), (see App.A) obtain relations on (4.27; 4.29). If the payoff volatility $\sigma^2(x)$ multiplied by factor $(1+2\alpha)$ is less than the mean payoff $x_0^2$ (4.30; A.5):

$$(1+2\alpha)\sigma^2(x) < {x_0}^2 \quad ; \quad \frac{1}{3} \leq \frac{1}{1+2\alpha} < 1 \tag{4.30}$$

Then (4.29) is always valid. If payoff volatility $\sigma^2(x)$ is high (A.6)

$$(1+2\alpha)\sigma^2(x) > \ {x_0}^2$$

Then (4.29) valid only for $\xi_{max}$ (A.6):

$$\xi_{max} < \frac{e_{t+1}[{x_0}^2+\sigma^2(x)]}{x_0\ [(1+2\alpha)\sigma^2(x)-{x_0}^2]}$$

However, this upper limit for $\xi_{max}$ can be high enough. The same, but more complex considerations can be presented for (3.2).

$$E_t[p^2 w''(c_t)] < -E[\beta x^2\ w''(c_{t+1})\ ]$$

## 5. Remarks on the Price Probability

Price probability is the major tool that defines the efficiency and accuracy of any financial forecast. However, the complexity of the economic origin of price probability is very securely hidden by agents' expectations, beliefs, and delusions. The hope for a quick and simple outcome is much more appealing than the discussion of the challenges that lie ahead. Let us consider that important issue with some more details. Below, we assume that all prices are adjusted to the current time *t.*

The conventional description of price probability "is based on the probabilistic approach and using A. N. Kolmogorov's axiomatic of probability theory, which is generally accepted now" (Shiryaev, 1999). The traditional definition of the price probability for the given time series is based on the frequency of trades at a price *p* during the averaging interval *Δ*. To describe price probability, it is enough to study *only a price time series* that presents *N* records, and *N* should be sufficiently large. Since Bachelier (1900), it has become standard to consider price time series as a random variable. "The probabilistic description of financial prices, pioneered by Bachelier." (Mandelbrot, et al., 1997). It is generally accepted that each of the *N* records of trades at price *p* during *Δ* has an equal probability ~ *1/N*. If there are *m(p)* trades at the price *p,* then the probability *P(p)* of the price *p* during *Δ* is estimated as *m(p)/N*. The use of the frequency of events is an absolutely correct and conventional description of a random price time series. The frequency-based approach to price probability checks how almost all standard probability measures (Christian Walck, 2007; Catherine Forbes, Merran Evans, Nicholas Hastings, and Brian Peacock, 2011) fit the description of the random market price. Parameters, which define standard probabilities, permit use them to increases plausibility and consistency with the observed random price time series. For different assets and markets, different standard probabilities are tested and applied to predict the random price dynamics as well as possible. Actually, the randomness of price $p(t_i)$ time series don't exist alone. Market price $p(t_i)$ is a result of a market deals at time $t_i$ and trivial trade price equation (5.1) describes the impact of trade value $C(t_i)$ and volume $U(t_i)$ on trade price $p(t_i)$:

$$C(t_i) = p(t_i)U(t_i) \tag{5.1}$$

The time series of prices are the results of many market transactions. The trade price equation (5.1) obliges us to consider the randomness of the market trade values $C(t_i)$ and volumes $U(t_i)$ as the economic origin of price stochasticity. The necessity of considering the impact of market trades on price randomness can be supported by a completely different issue. Indeed, as usual, investors assess the average price of shares in their portfolio simply as the ratio of the total value they spent to the total number of shares they purchased. Actually, there is almost no difference between the set of shares in the portfolio and the set of shares that were sold or purchased during $N$ market deals. One can consider the shares sold during $N$ transactions that were performed during the averaging interval $\Delta$ as a portfolio. Hence, to assess the average price of shares that were sold during the interval $\Delta$, one should use the same "portfolio assessment": to take the ratio of the total trade value to the total trade volume. That estimate of the average price exactly coincides with the definition of volume weighted average price (VWAP) that was introduced almost 35 years ago and is widely in use now (Stephen Berkowitz, Dennis Logue, Eugene Noser, 1988; Alexander Buryak and Ivan Guo, 2014; Enzo Busseti and Stephen Boyd, 2015; Darrell Duffie and Piotr Dworczak, 2018; CME Group, 2020). The definition of VWAP $p(t;1,1)$ is as follows: Assume that during $\Delta$ (5.5), there are $N$ market trades at moments $t_i$, $i=1,\ldots N$. Then VWAP $p(t;1,1)$ (5.2) that matches price equation (5.1) at time $t$ equals:

$$p(t;1,1) = \frac{1}{\sum_{i=1}^{N} U(t_i)} \sum_{i=1}^{N} p(t_i)U(t_i) = \frac{C_\Sigma(t;1)}{U_\Sigma(t;1)} = \frac{C(t;1)}{U(t;1)} \tag{5.2}$$

$$C_\Sigma(t;1) = \sum_{i=1}^{N} C(t_i) = \sum_{i=1}^{N} p(t_i)\,U(t_i) \quad ; \quad U_\Sigma(t;1) = \sum_{i=1}^{N} U(t_i) \tag{5.3}$$

$$C(t;1) = \frac{1}{N}\sum_{i=1}^{N} C(t_i) \quad ; \quad U(t;1) = \frac{1}{N}\sum_{i=1}^{N} U(t_i) \tag{5.4}$$

$$\Delta = \left[t - \frac{\Delta}{2}, t + \frac{\Delta}{2}\right] \quad ; \quad t_i \in \Delta\,, \; i = 1, \ldots N \tag{5.5}$$

We consider the time series of the trade value $C(t_i)$, volume $U(t_i)$ and price $p(t_i)$ as random variables during $\Delta$ (5.5). The relations (5.4) define the assessments of the average trade value $C(t;1)$ and average volume $U(t;1)$ by a finite number $N$ of market trades during $\Delta$ (5.5). The equation (5.1) at time $t_i$ defines the price $p(t_i)$ of a market transaction of value $C(t_i)$ and volume $U(t_i)$. The definition of VWAP $p(t;1,1)$ (5.2) results in the equation (5.6):

$$C(t;1) = p(t;1,1)\, U(t;1) \tag{5.6}$$

The equation (5.6) ties up the average trade value $C(t;1)$ and average volume $U(t;1)$ with the average trade price $p(t;1,1)$. However, for $m=2,3,\ldots$ the trade price equation (5.1) generates equations on the *m-th* power of trade value $C^m(t_i)$, volume $U^m(t_i)$ and price $p^m(t_i)$:

$$C^m(t_i) = p^m(t_i)U^m(t_i) \quad ; \quad m = 2,3\ldots \tag{5.7}$$

Similar to (5.1) and VWAP (5.2), equations (5.7) for each $m=2,3,..$ generate a set of average *n*-th power of price $p(t;n,m)$:

$$p(t;n,m) = \sum_{i=1}^{N} p^n(t_i)w(t_i;m) = \frac{1}{\sum_{i=1}^{N} U^m(t_i)} \sum_{i=1}^{N} p^n(t_i)U^m(t_i) \tag{5.8}$$

$$w(t_i;m) = \frac{U^m(t_i)}{\sum_{i=1}^{N} U^m(t_i)} \quad ; \quad \sum_{i=1}^{N} w(t_i;m) = 1 \tag{5.9}$$

For $n=m$

$$p(t;n,n) = \frac{1}{\sum_{i=1}^{N} U^n(t_i)} \sum_{i=1}^{N} p^n(t_i)U^n(t_i) = \frac{C_\Sigma(t;n)}{U_\Sigma(t;n)} = \frac{C(t;n)}{U(t;n)} \tag{5.10}$$

$$C_\Sigma(t;n) = \sum_{i=1}^{N} C^n(t_i) \quad ; \quad C(t;n) = \frac{1}{N}\sum_{i=1}^{N} C^n(t_i) \tag{5.11}$$

$$U_\Sigma(t;n) = \sum_{i=1}^{N} U^n(t_i) \quad ; \quad U(t;n) = \frac{1}{N}\sum_{i=1}^{N} U^n(t_i) \tag{5.12}$$

Relations (5.10; 5.11) describe the assessments of the *n-th* statistical moments of market trade values $C(t;n)$ and volumes $U(t;n)$ by a finite number $N$ of market trades. Similar to (5.6) one can use (5.8-5.11) to obtain equations on market trade *n-th* statistical moments:

$$C(t;n) = p(t;n,n)U(t;n) \tag{5.13}$$

Relations (5.2; 5.8-5.12) for the same time series of market price $p(t_i)$ give different estimates of price *n-th* statistical moments $p(t;n,m)$ (5.8) that are determined by different weighted functions $w(t_i;m)$ (5.9) related to different power $m$ of trade volume $U^m(t_i)$. We highlight that functions $w(t_i;m)$ (5.9) have the meaning of weighted functions but don't play role of price probability measures. For the case $U(t_i)$=const, $i=1,2,..N$ relations (5.2; 5.8; 5.10) define assessments of price statistical moments $p(t;n,n)$ that coincide with the frequency-based definition of price statistical moments (5.14):

$$p(t;n,n) = \frac{1}{N} \sum_{i=1}^{N} p^n(t_i) \tag{5.14}$$

As we mentioned above, the frequency-based price probability and price statistical moments (5.14) describe random properties of price $p(t_i)$ time series in the case that all trade volumes $U(t_i)$ are constant during the averaging interval $\Delta$ (5.5). That is not a good model of real markets.

It is well known that a random variable can equally be described by probability measure or by a set of statistical moments (Shiryaev, 1999). Equations (5.1; 5.7) cause, that market-based statistical moments of price must depend on statistical moments (5.11; 5.12) of trade value $C(t_i)$ and volume $U(t_i)$. To describe market-based price probability we should define its statistical moments. As the 1-st statistical moment $a(t;1)$ of the market-based price probability we take the VWAP $p(t;1,1)$ (5.2). Let us note $E_m[..]$ as mathematical expectation of the market-based price probability. Then we take:

$$E_m[p(t_i)] = a(t;1) = p(t;1,1) \tag{5.15}$$

To define the market-based price probability one should determine statistical moments $a(t;n)$:

$$a(t;n) = E_m[p^n(t_i)] \quad ; \quad n = 2,3,\ldots \tag{5.16}$$

However, one can't simply take price *n-th* statistical moments $p(t;n,n)$ (5.10) and define market-based statistical moments $a(t;n)$ (5.16). For different $n$, the statistical moments $p(t;n,n)$ are determined by different weight functions $w(t;n)$, and hence, some $p(t;n,n)$ (5.10) can be incompatible with others. For example, $p(t;2,2)$ could be less than $p^2(t;1,1)$. If so, the usage of $p(t;2,2)$ as the 2-d market-based statistical moment $a(t;2)$ could result in negative price volatility, and that does not make any sense. To avoid such mistakes, one should require even central statistical moments to be non-negative. To define the 2-d market-based statistical moment $a(t;2)$, we set:

$$E_m[\delta^2 p(t,t_i)] = \frac{1}{\sum_{i=1}^{N} U^2(t_i)} \sum_{i=1}^{N} \delta^2 p(t,t_i) U^2(t_i) = \sum_{i=1}^{N} \delta^2 p(t,t_i) w(t_i;2) \tag{5.17}$$

$$\delta p(t,t_i) = p(t_i) - a(t;1) \tag{5.18}$$

Relations (5.17) describe the market-based central 2-d statistical moment $E_m[\delta^2 p]$ (5.17; 5.18) to be equal the central 2-d moment determined by weighted function $w(t_i;2)$ (5.9). Relations (5.17; 5.18; 5.10) define the *2-d* statistical moment $a(t;2)$ (5.16) as follows:

$$E_m[\delta^2 p(t,t_i)] = a(t;2) - a^2(t;1) = \sigma^2(t) \tag{5.19}$$

$$\frac{1}{\sum_{i=1}^{N} U^2(t_i)} \sum_{i=1}^{N} \delta^2 p(t,t_i) U^2(t_i) = p(t;2,2) - 2p(t;1,2)a(t;1) + a^2(t;1) \tag{5.20}$$

From (5.19; 5.20) obtain market-based price 2-d statistical moment $a(t;2)$:

$$a(t;2) = p(t;2,2) + 2a(t;1)[a(t;1) - p(t;1,2)] \tag{5.21}$$

Price volatility $\sigma^2(t)$ (5.19) takes form (5.22):

$$\sigma^2(t) = p(t;2,2) - 2p(t;1,2)a(t;1) + a^2(t;1) \tag{5.22}$$

Due to (5.17; 5.20), price volatility $\sigma^2(t)$ (5.22) is always non-negative. We highlight that market-based price statistical moment $a(t;1)$ *(5.15)*, $a(t;2)$ (5.21) and price volatility $\sigma^2(t)$ (5.22) depend on price statistical moments $p(t;1,1)$ (5.2), $p(t;2,2)$, $p(t;1,2)$ (5.8) determined by weighed functions $w(t_i;1)$ and $w(t_i;2)$ (5.9). The same time, the 2-d market-based price statistical moment *a(t;2)* (5.21) and price volatility $\sigma^2(t)$ (5.22) depend on 1-st and 2-d statistical moments of market trade value *C(t;1), C(t;2)* (5.11) and volume *U(t;1), U(t;2)* (5.12). To predict the market-based price volatility $\sigma^2(t)$ (5.22) at horizon *T,* one must forecast the market trade statistical moments *C(t;1), C(t;2)* (5.11), and *U(t;1), U(t;2)* (5.12) at the same horizon *T.* In simple words, to predict price volatility, one should forecast the average and volatilities of market trade value $\sigma_C^2(t)$ and trade volume $\sigma_U^2(t)$ (5.23):

$$\sigma_C^2(t) = C(t;2) - C^2(t;1) \quad ; \quad \sigma_U^2(t) = U(t;2) - U^2(t;1) \tag{5.23}$$

For brevity, in this paper we reduce the description of the market-based price probability by the first two price statistical moments *a(t;1)* (5.15) and *a(t;2)* (5.21), and shall consider the market-based price statistical moments in more detail in further publications. Now we describe how the introduction of market-based statistical moments *a(t;1)* and *a(t;2)* impacts the long-studied problem of price-volume correlations.

## 6. Price-Volume Correlations

Correlations between two random variables are determined by their probability measures. The introduction of market-based price statistical moments *a(t;1)* (5.15) and *a(t;2)* (5.21) allows us to derive the form of the price-volume correlations. We take *a(t;1)* (5.15) to be equal to VWAP *p(t;1,1)* (5.2), and that results in zero correlations between the time series of the trade volume $U(t_i)$ and price $p(t_i)$ during the averaging interval $\Delta$ (5.5). From (5.1; 5.2; 5.4), obtain:

$$E[p(t_i)U(t_i)] = E[C(t_i)] = \frac{1}{N}\sum_{i=1}^{N} C(t_i) = \frac{1}{N}\sum_{i=1}^{N} p(t_i)U(t_i) =$$

$$= \frac{1}{\sum_{i=1}^{N} U(t_i)}\sum_{i=1}^{N} p(t_i)U(t_i) \cdot \frac{1}{N}\sum_{i=1}^{N} U(t_i) = E_m[p(t_i)]E[U(t_i)] \tag{6.1}$$

Hence, from (6.1), obtain that the correlation $corr\{p(t_i)U(t_i)\}$ (6.2) between the time series of price $p(t_i)$ and trade volume $U(t_i)$ equals zero:

$$corr\{p(t_i)U(t_i)\} \equiv E[p(t_i)U(t_i)] - E_m[p(t_i)]E[U(t_i)] = 0 \tag{6.2}$$

The correlation $corr\{p(t_i)U^2(t_i)\}$ between price $p(t_i)$ and squares of volumes $U^2(t_i)$ takes form:

$$E[p(t_i)U^2(t_i)] = E[C(t_i)U(t_i)] = E[C(t_i)]E[U(t_i)] + corr\{C(t_i)U(t_i)\}$$

$$corr\{p(t_i)U^2(t_i)\} = E[C(t_i)U(t_i)] - a(t;1)U(t;2)$$

$$E[C(t_i)U(t_i)] \sim \frac{1}{N}\sum_{i=1}^{N} C(t_i)\,U(t_i)$$

From above and (5.23), obtain:

$$corr\{p(t_i)U^2(t_i)\} = corr\{C(t_i)U(t_i)\} - a(t;1)\sigma_U^2(t) \tag{6.3)}$$

The correlation $corr\{p^2(t_i)U^2(t_i)\}$.

$$corr\{p^2(t_i)U^2(t_i)\} = E[p^2(t_i)U^2(t_i)] - a(t;2)U(t;2)$$

$$E[p^2(t_i)U^2(t_i)] = E[C^2(t_i)] = C(t;2)$$

From (5.13; 5.21), obtain:

$$corr\{p^2(t_i)U^2(t_i)\} = 2a(t;1)U(t;2)[p(t;1,2) - a(t;1)] \quad (6.4)$$

Actually, many publications describe positive or negative correlations between price and trading volume (George Tauchen and Mark Pitts, 1983; Jonathan Karpoff, 1987; John Campbell, Sanford Grossman, and Jiang Wang, 1993; Guillermo Llorente, Roni Michaely, Gideon Saar, and Jiang Wang, 2001; Anthony DeFusco, Charl Nathansona, and Eric Zwick, 2017). In these papers, authors use the frequency-based price probability (5.14) to estimate correlations $corr\{p(t_i)U(t_i)\}$. The use of different probabilities causes different results in assessments of correlations. The use of market-based price statistical moments and VWAP reveals no correlations between trade volume and price $corr\{p(t_i)U(t_i)\}=0$ (6.2), and allows derive expressions for $corr\{p(t_i)U^2(t_i)\}$ (6.3) and $corr\{p^2(t_i)U^2(t_i)\}$ (6.4).

## 7. Conclusion

Each economic theory and asset pricing in particular should directly indicate the time scales $\Delta$ of the proposed approximation. The time series of the market trades with time shift $\varepsilon$ introduces the initial division of the time axis. Asset pricing models should take into account these initial data as the only source for averaging market time series that presumes the usage of a particular averaging interval $\Delta>>\varepsilon$. Averaging the initial market time series during $\Delta$ introduces a transition from the initial time axis division multiple of $\varepsilon$ to the new time division multiple of $\Delta$. To consider the utility function and price "today" and "next day," one should use the same time axis division "today" and "next day" and hence use the same averaging interval $\Delta$. Averaging the investor's utility function "today" and "next day" introduces modifications to the investor's utility and basic pricing equation. We show that the conventional basic pricing equation (2.1):

$$p = E[m\,x] \quad (7.1)$$

should be complemented by the modified basic pricing equation (3.3):

$$E_t[p\,w'(c_t)] = \beta E\,[x\,w'(c_{t+1})] \quad (7.2)$$

that takes into account the averaging procedures "today" and "next day" during the same averaging interval. The choice of interval $\Delta$ allows considering the Taylor series expansions of the modified basic pricing equation (3.3; 7.2) by price and payoff fluctuations and subsequent averaging of fluctuations. For linear and quadratic approximations of the modified basic pricing equation, we obtain the average price, price volatility, mean payoff, payoff volatility, etc. In the linear approximation, (4.12) presents the dependence of the mean price $p_0$ "today":

$$p_0 = m_0 x_0 + \xi_{max}[m_1 \sigma^2(x) + m_2 \sigma^2(p)] \quad (7.3)$$

on price volatility $\sigma^2(p)$ "today" and on the mean payoff $x_0$, payoff volatility $\sigma^2(x)$ "next day" and on the asset's amount $\xi_{max}$ that defines the max of the utility and equals the root of the equation (3.3). On the one hand, (7.3) modifies the conventional statement "price equals expected discounted payoff" and demonstrates dependence on price volatility $\sigma^2(p)$ "today". We highlight that (7.3) uncovers the direct dependence of the mean price $p_0$ "today" on the asset's amount $\xi_{max}$. That direct dependence doesn't add confidence in the impeccability of consumption-based and similar pricing models, and further argumentation is required to solve the troubles that arise with the direct dependence of mean price $p_0$ (7.3) on $\xi_{max}$.

We consider the description of market-based price probability and statistical moments as the core issues of any pricing or financial model. We used VWAP as the 1-st market-based statistical moment and introduced the 2-d market-based price statistical moment and the price volatility. That highlights the direct dependence of market-based statistical moments on the randomness of market trade. The predictions of the average price and price volatility are determined by the forecasts of market trade statistical moments.

The definition of the 1-st statistical moment as VWAP results in zero correlations between price and trade volume time series $corr\{p(t_i)U(t_i)\}=0$ (6.2), but we derive non-zero correlations $corr\{p(t_i)U^2(t_i)\}$ (6.3) and $corr\{p^2(t_i)U^2(t_i)\}$ (6.4).

This trinity – the averaging interval $\Delta$, the Taylor series, and the market-based price probability - can provide successive approximations for other versions of asset pricing, financial, and economic models.

## Appendix A.

## Utility Maximum

We start with (4.29):

$$p^2 > -\beta \frac{w''(c_{t+1;0})}{w''(c_t)} [{x_0}^2 + \sigma^2(x)] - \beta \frac{w'''(c_{t+1;0})}{w''(c_t)} \xi_{max} [2x_0\sigma^2(x) + \gamma^3(x)] \qquad \text{(A.1)}$$

If the right side is negative, then it is always valid. If the right side is positive, then there exists a lower limit on the price $p$. For simplicity, neglect term $\gamma^3(x)$ to compare with $2x_0\sigma^2(x)$ and take the conventional power utility $w(c)$ (Cochrane, 2001) as:

$$w(c) = \frac{1}{1-\alpha} c^{1-\alpha} \qquad \text{(A.2)}$$

Let us consider the case with the negative right side (A.1). Simple but long calculations give:

$$-\beta \frac{w''(c_{t+1;0})}{w''(c_t)} [{x_0}^2 + \sigma^2(x)] < \beta \frac{w'''(c_{t+1;0})}{w''(c_t)} \xi_{max}\, 2x_0\sigma^2(x)$$

$$\xi_{max}\, 2x_0\sigma^2(x) < -\frac{w'''(c_{t+1;0})}{w''(c_{t+1;0})} [{x_0}^2 + \sigma^2(x)] \qquad \text{(A.3)}$$

Let us take into account (A.2) and for (A.3) obtain:

$$\frac{w''(c)}{w'''(c)} = -\frac{c}{1+\alpha} \quad ; \quad \xi_{max}\, 2x_0\sigma^2(x) < \frac{e_{t+1} + x_0\xi_{max}}{1+\alpha} [{x_0}^2 + \sigma^2(x)]$$

$$\xi_{max} x_0 [(1+2\alpha)\sigma^2(x) - {x_0}^2] < e_{t+1}[{x_0}^2 + \sigma^2(x)] \qquad \text{(A.4)}$$

Inequality (A.4) determines that the right side (A.1) is negative in two cases. The left side of (A.4) is negative and

$$(1+2\alpha)\sigma^2(x) < {x_0}^2 \quad ; \quad \frac{1}{3} \le \frac{1}{1+2\alpha} < 1 \qquad \text{(A.5)}$$

Inequality (A.5) describes small payoff volatility $\sigma^2(x)$. In this case, the right side of (A.1) is negative for all $\xi_{max}$ and all price $p$ and hence (4.27) that defines the max of utility (2.5) is valid. The left side of (A.4) is positive, and

$$(1+2\alpha)\sigma^2(x) > {x_0}^2 \quad ; \quad \xi_{max} < \frac{e_{t+1}[{x_0}^2 + \sigma^2(x)]}{x_0\, [(1+2\alpha)\sigma^2(x) - {x_0}^2]} \qquad \text{(A.6)}$$

This case describes high payoff volatility and the upper limit on $\xi_{max}$ to utility (2.5). Take the positive right side in (A.1). Then (A.4) is replaced by the opposite inequality

$$\xi_{max} x_0 [(1+2\alpha)\sigma^2(x) - {x_0}^2] > e_{t+1}[{x_0}^2 + \sigma^2(x)] \qquad \text{(A.7)}$$

It is valid for (A.6) only. (A.7) determines a lower limit on $\xi_{max}$ to utility (2.5):

$$\xi_{max} > \frac{e_{t+1}[{x_0}^2 + \sigma^2(x)]}{x_0\, [(1+2\alpha)\sigma^2(x) - {x_0}^2]}$$